\documentstyle[preprint,aps]{revtex}
\begin{document}
\draft
\tightenlines
\title{Cosmological constant and spontaneous gauge symmetry breaking: 
the  particle physics and cosmology interface charade}
\author{F. Pisano and N. O. Reis}
\address{Department of Physics, Federal University of Paran\'a, 
         Curitiba, PR 81531-990, Brazil}
\date{\today}
\maketitle
\begin{abstract}
We describe one of the remarkable problems of theoretical physics 
persevering up to the beginning of the millennium. All gauge theories with 
spontaneous gauge symmetry breaking from the standard model of particle 
physics with the electroweak symmetry breaking at the Fermi scale, 246 GeV, 
up to strings, supergravity, 
and the M(embrane)-theory superunification 
with symmetry breaking starting near the Planck 
scale, $10^{19}$ GeV, foresee that the spontaneous symmetry breakings 
induce a vacuum energy at least $50$ orders of magnitude 
larger than the stringent experimental bound $G\Lambda\lesssim 10^{-122}$ 
on the value of 
the cosmological constant $\Lambda$. This fact seems to 
have a universal character since it occurs from the Fermi scale up to the 
Planck one. It is the vacuum catastrophe.
\end{abstract}
\section{Introduction}
According to the general relativity~\cite{AE,MTW}, the vacuum energy density 
has a defined meaning since it couples unavoidably 
with gravitation and can be 
parametrized by a magnitude known as cosmological constant, $\Lambda$. 
The cosmological bounds on $\Lambda$ are severe and as we will see they 
result in the smallest number of physics directly related to the 
fundamental universal 
interactions. In the year of 1922, A.A. Friedmann~\cite{AAF22} 
(1888--1925) demonstrated that the Einstein's general relativity equations 
admit non-static solutions connected to an expanding Universe. In a first 
stage, Einstein even demonstrated that the argument conducing to the 
conclusion of Friedmann is wrong~\cite{AE22}. Nevertheless, he found out 
his own mistake~\cite{AE23} and started to consider the Friedmann's 
theoretical results as clarifying. With the discovery of the expansion of 
the Universe by Edwin P. Hubble~\cite{EPH29} (1889--1953), according to 
whom there is a linear relation between velocities and distances at 
cosmological scale, Einstein 
disregarded definitely the cosmological constant and confirmed besides 
W. de Sitter~\cite{AEdS32} (1872--1934) that this term is completely 
unsatisfactory theoretically. 
\par
The standard model of the non-gravitational interactions~\cite{GWS} with 
the internal symmetry gauge semisimple group
\begin{equation}
{\cal G}_{321}\equiv 
{\rm SU(3)}_c\otimes {\rm SU(2)}_L\otimes {\rm U(1)}_Y
\label{uno}
\end{equation}
has resisted to the experimental challenges~\cite{JLR} being proved in 
tests of accuracy with great success~\cite{PL95}. It was also verified 
the experimental indication of incompleteness of the standard model 
related to the fermionic nature of the three neutrino flavors. It has 
been noticed that there is oscillation among the different neutrino 
flavors~\cite{SK} which can happen whether the fermion is described by a 
four components Dirac~\cite{PAMD} spinor or by the two components 
Majorana~\cite{EM37} state and not only by a pair of Weyl~\cite{HW19} 
eigenstates of chirality,
\begin{equation}
\gamma_5\psi = -\psi,\quad \bar\psi\gamma_5 = \bar\psi
\label{ddue}
\end{equation}
where $\gamma_5=\gamma^5={\rm i}\gamma^0\gamma^1\gamma^2\gamma^3$, 
$ {\rm i}={\sqrt{-1}}$, and $\gamma^\mu$, $\mu=0,1,2,3$  
are the Dirac matrices. The chiral invariance 
does not reveal itself since any fundamental fermion is massive and a Dirac 
mass term $m_\psi\bar\psi\psi$ violates the chiral invariance due to an 
algebraic sign change. The good symmetry to provide the dynamics is the 
{\it local} gauge symmetry related to the properties of the gauge bosons 
which transmits the interactions instead of the {\it global} symmetries 
such as the U(1) symmetry associated with the conservation of lepton 
number or with the baryon charge. Nevertheless, the electromagnetic U(1) 
local symmetry, associated with the conservation of the electric charge, is 
a local gauge symmetry and not a global one, whose group dimension is one 
and thus there is a gauge boson associated, the photon, and as the group 
rank is also one this gauge boson is electrically neutral and is also its 
own antiparticle. 
\par
The origin of the unified description of the dynamics of the fundamental 
interactions retrace to the pioneer attempts; in 1914 of 
G. Nordstr\"om~\cite{GN14} (1881--1923) and, soon later, still in the 
twenties, of T. Kaluza~\cite{TK21} (1885--1954) and O. Klein~\cite{OK26} 
(1849--1925). Originally it was proposed the unification of the Maxwell 
theory of the electrodynamics with the general relativity in a 
space-time of five dimensions. The fifth dimension would be compactified 
in the scale of gravity quantization which is the Planck scale, 
$\sim 10^{19}\,\,{\rm GeV}\simeq 10^{-35}\,\,{\rm m}$, 
where $1\,\,{\rm GeV}\,\,({\rm gigaelectron\,\,volt})
\equiv 10^9\,\,{\rm eV}$ with 
$1\,\,{\rm eV}=1.602\times 10^{-19}\,\,{\rm J}$, being therefore 
undetectable nowadays. The fifth dimension $x_5$ has a circle topology, 
\begin{equation}
\phi (x_5,r) = \phi (x_5 + 2\pi r),
\label{ttre}
\end{equation}
where $r\simeq 10^{-35}$ m. The non-trivial generalization of the 
space-time Kaluza--Klein extension to the space of internal symmetries 
result in the Yang--Mills~\cite{CNYRM} theories built up in a space with 
$((3+1)+N)$ dimensions decomposed as the product of the Minkowski flat 
space-time, $M_{(3+1)}$, with a manifold $G$ of dimension $N$ and so 
$M_{(3+1)}\otimes G$ is the complete gauge group. This is the base of the 
non-Abelian gauge theories which describe all interactions including the 
gravitation since the general relativity is a gauge theory par excellence 
in which the internal symmetry transformations correspond to the 
general transformations of coordinates. The masses of all the fermions 
including the neutrinos~\cite{SK}, the gauge bosons, and also of the 
Higgs scalar boson, the particle undetected so far, 
are generated in the process of spontaneous breaking of gauge symmetry by 
the Higgs--Kibble~\cite{Higgsssb,Kibble} mechanism. Always that a 
spontaneous breaking of the gauge symmetry occurs in the theory, from the 
standard model of the elementary particles in the Fermi scale, 
$\simeq 246$ GeV, up to superstrings~\cite{LNZ} in the Planck scale, 
$\simeq 10^{19}$ GeV, it is generated a constant term in the 
scalar potential that corresponds to a great increase in the 
energy density of the vacuum 
state. In the realm of the study of the high energy phenomenological 
processes it is even possible to simply dismiss this term. However, this 
constant term contributes to the value of the cosmological constant as we 
could see ahead. 
\par
The general relativity~\cite{AE15} is formulated as a classical 
field theory. 
All of the attempts of quantization have always resulted 
in non-renormalizable theories~\cite{West}. 
The purpose is to accomplish the unification of the gravitation with the 
other interactions so that the infinities appearing in the different 
sectors cancel themselves order-by-order in the perturbative serie, 
resulting in a renormalizable theory. The same theory would offer the reason 
why the cosmological constant $\Lambda$ is so small, 
\begin{equation}
G\Lambda \lesssim 10^{-122}
\label{quatt}
\end{equation}
where $G=6.673(10)\times 10^{-11}$ kg$^{-1}$ m$^3$ sec$^{-2}$ 
is the Newton--Cavendish universal constant of gravitation. 
At present, the most successful of the candidates to accomplish this 
unification is the local supersymmetry~\cite{VoAk} containing the gravity, 
being known as supergravity~\cite{FFer} which is one of the facts of the 
superstrings~\cite{DPh} theories. Today, the partial, grand, and the 
complete or total unification possibilities 
are: compositeness, technicolor, grand unification, symmetric 
left-right gauge groups, chiral gauge groups, 
Kaluza--Klein models, supersymmetry, 
supersymmetric grand unification, supergravity, superstrings~\cite{Green} 
and the superunification Membrane-theory~\cite{WittenPT} 
with a total unification of all fundamental interactions.
\section{One century ago: Ultraviolet Catastrophe}
In June of 1900, Lord Rayleigh~\cite{LRa} (J.W. Strutt, 1842--1919) 
suggested, for the first time, the application to the thermal bath of a 
black body of the principle of equipartition of energy by the 
Maxwell--Boltzmann distribution,
\begin{equation}
f(v,T) = \frac{{\rm d}N}{{\rm d}v}\propto 
v^2\exp\{-\frac{mv^2}{2}\frac{1}{kT}\}
\label{ccinn}
\end{equation}
which is the statistical distribution of an ensemble with $N$ 
indistinguishable particles of any kind with mass $m$ of a system in 
equilibrium at a temperature $T$, where 
$k=1.381\times 10^{-23}\,\,{\rm J}\,{\rm K}^{-1}$ is the Boltzmann 
constant. The function $f(v,T)$ will be zero when $v=0$, will reach the 
maximum value when $v=(2kT/m)^{\frac{1}{2}}$ and will reduce rapidly until 
zero with the ulterior increase of the velocity. The Maxwell--Boltzmann 
distribution has a universal character but it is founded on the classical 
physics so that its application could not be appropriate to quantum systems 
of fermions or bosons. Actually, the classical distribution of 
Maxwell--Boltzmann when applied to a cavity of harmonic oscillators does 
not provide the correct energy density but 
\begin{equation}
\rho (\nu,T)=\frac{8\pi}{c^3}\nu^2\,kT
\label{ssei}
\end{equation}
which is a function of the square of the frequency. This turned to be 
recognized as the Rayleigh--Jeans formula. The $\nu^2T$ law was also 
obtained by Lorentz~\cite{HL1903} and Einstein~\cite{AE1905} in 1905. 
The divergent behavior of $\rho (\nu,T)$ for high frequencies was named 
by Paul Ehrenfest (1880--1933) as `ultraviolet catastrophe' in 
1911~\cite{Milloni}. Lord Rayleigh, in his work of 1900~\cite{LRa} does not 
calculate the constant
\begin{equation}
c_1 = \frac{8\pi}{c^3}k
\label{sette}
\end{equation}
in 
\begin{equation}
\rho (\nu,T) = c_1\nu^2 T
\label{ootto}
\end{equation}
but in order to supply the catastrophical divergent behavior at high 
frequencies he introduced {\it ad hoc} a cutoff exponential factor 
thereby suggesting the radiation law
\begin{equation}
\rho (\nu,T) = c_1\nu^2 T\exp\{c_2\nu/T\}
\label{nnovv}
\end{equation}
known as Rayleigh law. In the year of 1905 he retake his $\nu^2T$ law, 
then determining the $c_1$ constant but obtaining $c_1/8$. The mistake is 
corrected by Sir James Hopwood Jeans~\cite{JHJ} (1887--1946) who is 
thanked by Lord Rayleigh~\cite{WSRay} by his contribution. 
\par
It is not known whether M. Planck~\cite{MKERP} (1858--1947) knew of the 
June work of Lord Rayleigh. Anyhow he does not mention such article, 
neither does Lorentz~\cite{HL1903}. 
However it is worthy to notice that Planck made reference to the inspiration 
he had received from the statistical methods of 
L. Boltzmann~\cite{APais} (1844--1916). The job accomplished by Planck was 
to begin with the Maxwell--Boltzmann classical distribution but treating 
the energy content of the stationary electromagnetic waves in the cavity of 
a black body as a discrete magnitude replacing the integral continuous 
summation by a discrete sum indexed with the principal quantum number in 
the energy density distribution. The Planck job is well known and is called 
{\it quantization}. Immediately, he got the spectral density
\begin{equation}
\rho (\nu,T) = \frac{8\pi}{c^3}\nu^2\frac{h\nu}{\exp\{h\nu/kT\}-1}
\label{diecc}
\end{equation}
where $h=6.626\,068\,76(52)\times 10^{-34}\,\,{\rm J}\,{\rm sec}$ is 
the Planck constant~\cite{PDG2000}. Under the condition that $h\nu/kT\gg 1$ 
the Planck distribution fall back in the W. Wien~\cite{Wien} (1864--1928) law 
from 1896,
\begin{equation}
\rho (\nu,T) = \frac{8\pi}{c^3}\nu^2\,[h\nu\exp\{-h\nu/kT\}]
\label{unndd}
\end{equation}
which is also correct in the quantum realm, $h\nu/kT\gg 1$, as constated 
for the first time by the remarkable experiment of F. Pashen~\cite{Pashen} 
(1865--1947) in which $h\nu/kT\simeq 15$ for $T=10^3$ K and 
$\lambda = 1\mu{\rm m} = 10^{-6}$ m. 
\par
In 1906, Einstein~\cite{AE1906} realized that the Planck theory of 1900 
uses implicitly the hypothesis of the quantum of radiation that is a 
quantum property of free electromagnetic radiation. The energy of an 
oscillator of the black body electromagnetic thermal bath could take only 
values which are integer multiples of $h\nu$. In the processes of emission 
and absorption the energy of an oscillator varies only by integer multiples 
$nh\nu$ of the quantum $h\nu$, being $n$ the principal quantum number 
related to the quantization of energy. Hence, the photon is a state of the 
electromagnetic field with a frequency $\nu$ and a wave vector ${\bf k}$ well 
defined linked, respectively, to the energy
\begin{equation}
E = h\nu
\label{dodci}
\end{equation}
and to the linear momentum
\begin{equation}
{\bf p} = h {\bf k}
\label{trdci}
\end{equation}
which satisfies the relativistic equation of energy
\begin{equation}
E = c|{\bf p}|
\label{quatrdici}
\end{equation}
for a massless particle. The word {\it photon} appeared for the first time 
in an article of 1926 by Gilbert Lewis~\cite{Lewis}.
\par
It is interesting to notice that both Planck and Einstein and also Lorentz 
presented serious restrictions in relation to the quantum of action which 
retains its individuality in the propagation~\cite{APais}. Einstein, in a 
letter~\cite{APais} of 1951, wrote `All these fifty years of meditation have 
not approached me to the answer to the question: {\it What are the quanta of 
light?}' 
\section{The Standard Model of non-gravitational fundamental interactions}
The standard model of the non-gravitational 
fundamental interactions, which are the two 
nuclear ones, 
the `weak' and the `strong' interactions, and the electromagnetic 
interaction consists of three relativistic 
quantum gauge field theories with the corresponding gauge symmetries 
${\rm SU(3)}_c$ and ${\rm SU}(2)_L\times${\rm U}(1)$_Y$ 
associated to the 
quantum chromodynamics~\cite{GWS} (QCD) and, in the electroweak sector 
SU(2)$_L\otimes$U(1)$_Y$, to the quantum flavor dynamics~\cite{GWS} (QFD) 
and to the quantum electrodynamics~\cite{QED} (QED), the only one among the 
three ones which is an Abelian theory. The gauge coupling constants of the 
electroweak sector, $g$ of SU(2)$_L$ and $g^\prime$ of U(1)$_Y$, and the 
parameter $\theta_{\rm W}$, $\tan\theta_{\rm W}=g^\prime/g$, are all free 
parameters, so that there is no (grand) unification of the interactions 
even in the called electroweak sector. The c-number $Y$ that indexes the 
Abelian factor U(1)$_Y$ is the weak hypercharge related to the electric 
charge operator
\begin{equation}
\frac{{\cal Q}}{|e|} = T_3 + T_0
\label{quindci}
\end{equation}
where $T_3=\frac{1}{2}{\rm diag}(+1,-1)$ is one half of the third Pauli 
traceless matrix and $T_0=Y\frac{1}{2}{\rm diag}(+1,+1)$. 
\par
The fundamental chiral fermions are grouped in three generations of leptons 
and quarks  which are the fundamental constituents of all matter. To any 
fermion $\psi$  we define its chiral left-handed ($L$) component
\begin{mathletters}
\begin{equation}
\psi_L = P_L\psi
\label{sedici}
\end{equation}
and the right-handed ($R$) one
\begin{equation}
\psi_R = P_R\psi
\label{diciaset}
\end{equation}
\end{mathletters}
where 
\begin{mathletters}
\begin{equation}
P_L = \frac{1}{2}(1-\gamma_5),
\label{diciotto}
\end{equation}
\begin{equation}P_R = \frac{1}{2}(1+\gamma_5)
\label{dicnove}
\end{equation}
\end{mathletters}
are the chiral idempotent projectors. 
Counting three SU(3)$_c$ color charges for each quark flavor each generation 
contains 15 Weyl fermions in two chiral states, $L$ and $R$, which are 
attributed to the 
{\it fundamental representation} 
of the ${\cal G}_{321}$ gauge 
group. For three families of fermions there are 45 massless Weyl fermions 
which under the gauge group are attributed to the following representations:\\
Chiral left-handed leptons transform under ${\cal G}_{321}$ as
\begin{mathletters}
\begin{equation}
\Psi_{\ell L} = \left (
\begin{array}{c}
\nu_\ell \\
\ell
\end{array}
\right)_L\sim ({\bf 1}_c,{\bf 2}_L,Y=-1)
\label{venti}
\end{equation}
while chiral right-handed components transform as singlets under all 
the factors of the gauge semisimple group,
\begin{equation}
\ell_R\sim ({\bf 1}_c,{\bf 1}_R,Y=-2)
\label{ventuno}
\end{equation}
\end{mathletters}
for the $\ell = e^-,\mu^-,\tau^-$ flavors. Three families of left-handed 
chiral quarks are attributed to color triplets but also to SU(2)$_L$ flavor 
doublets
\begin{mathletters}
\begin{equation}
\Psi_{qL} = \left (
\begin{array}{c}
{\cal U}_q \\
{\cal D}_q
\end{array}
\right )_L\sim ({\bf 3}_c,{\bf 2}_L,Y=1/3)
\label{vendue}
\end{equation}
with the respective flavor $R$-singlets
\begin{equation}
{\cal U}_{qR}=\{u_{qR}, c_{qR}, t_{qR}\}
\sim ({\bf 3}_c,{\bf 1}_R,Y=4/3)
\label{ventre}
\end{equation}
\begin{equation}
{\cal D}_{qR}=\{d_{qR}, s_{qR}, b_{qR}\}
\sim
({\bf 3}_c,{\bf 1}_R,Y=-2/3)
\label{venquat}
\end{equation}
\end{mathletters}
where $q=1,2,3$ denotes the SU(3)$_c$ color index 
({\it red}, {\it green}, {\it blue}) and 
${\cal U}_{qR}$ and ${\cal D}_{qR}$ denote the three flavors with 
electric charges $\pm\frac{2}{3}|e|$ and more three corresponding flavors 
with the electric charges $\mp\frac{1}{3}|e|$ for the particle and 
antiparticle states. It is contained in the inner of a same family of quarks 
12 Weyl two-component spinors corresponding to two flavors, 
type `up,' ${\cal U}_q$, and `down,' ${\cal D}_q$, 
with two states of chirality, $L$ and $R$, and still three flavor 
degrees of freedom. Therefore, counting the three spinors of the leptonic 
sector, it results in an ammount of 15 Weyl spinors. 
\par
Otherwise, the entire sector of gauge bosons is attributed to the {\it adjoint 
representation} of the ${\cal G}_{321}$ group. The color factor, SU(3)$_c$, 
with a dimension $3^2-1=8$, contains eight generators which are the 
$3\times 3$ Gell-Mann matrices closing among them the Lie algebra
\begin{equation}
[\lambda_a,\lambda_b]=2{\rm i} f_{abc}\lambda_c,\quad a=1,2,...,8
\label{vecinq}
\end{equation}
associated to eight gauge bosons, the gluons $g^a_\mu$, which are the 
transporter of the strong nuclear interaction among hadrons, and also to 
eight non-vanishing group structure constants $f_{abc}$. The rank of the 
color group is $3-1=2$, corresponding to the number of generators that 
are diagonal in the matricial representation. Likewise, in the case of the 
SU(2)$_L$ flavor group the dimension $2^2-1=3$ correspond to the three 
group generators which are one half times the $\sigma_k$ Pauli matrices 
satisfying the Lie algebra
\begin{equation}
[\sigma_k,\sigma_l]=2{\rm i}\epsilon_{klm}\sigma_m,\quad k=1,2,3
\label{vensei}
\end{equation}
and to the symmetry eigenstates gauge bosons, $W^k_\mu$, and, finally, to 
the unique generator, $Y/2$, of the U(1)$_Y$ Abelian factor with the gauge 
boson $B_\mu$. The mass eigenstates physical states are a pair of electrically 
charged gauge bosons
\begin{mathletters}
\begin{equation}
W^\pm_\mu = \frac{1}{\sqrt 2}(W^1_\mu\pm {\rm i}W^2_\mu)
\label{vesette}
\end{equation}
with mass~\cite{PDG2000} $80.419(56)$ GeV$/c^2$ and the neutral gauge bosons, 
the photon,
\begin{equation}
A_\mu = \cos\theta_{\rm W} W^3_\mu+\sin\theta_{\rm W} B_\mu
\label{ventoto}
\end{equation}
and 
\begin{equation}
Z_\mu = -\sin\theta_{\rm W} W^3_\mu+\cos\theta_{\rm W} B_\mu
\label{venove}
\end{equation}
\end{mathletters}
with mass~\cite{PDG2000} $91.1882(22)$ GeV$/c^2$ in which, as masses, the 
parameter $\sin^2\theta_{\rm W}=0.23147(16)$ is one of the twenty free 
parameters of the standard model. Almost a half of the free parameters of 
the whole standard model are masses. 
\par
It can be realized now that each fermion family contains four chiral quarks. 
Nonetheless, the leptons of each family consist only of three Weyl spinors, 
$\ell_L$, $\ell_R$, and $\nu_{\ell L}$. What is the reason of this 
asymmetry between leptons and quarks? This difference is the reason of the 
neutrino physics and of the fact that the electroweak sector 
SU(2)$\otimes$U(1) to have the chiral character 
${\rm SU}(2)={\rm SU}(2)_L$ and not the symmetric one~\cite{WPSetal} 
${\rm SU}(2)={\rm SU}(2)_L\otimes {\rm SU}(2)_R$. Oscillations of neutrino 
flavors were confirmed~\cite{SK}, indicating that they could be massive 
fermions so that the independent Weyl pairs
$(\nu_{\ell L}, \bar\nu^c_{\ell R})$ and 
$(\nu_{\ell R}, \bar\nu^c_{\ell L})$ involving the charge conjugation 
operation,
\begin{mathletters} 
\begin{equation}
(\nu_{\ell L})^c = (P_L\nu_\ell)^c = 
\frac{1}{2}(1+\gamma_5)\gamma^0\nu^*_\ell = (\nu^c_\ell)_R\,,
\label{trenta}
\end{equation}
and
\begin{equation}
(\nu_{\ell R})^c = (P_R\nu_\ell)^c = 
\frac{1}{2}(1-\gamma_5)\gamma^0\nu^*_\ell = (\nu^c_\ell)_L
\label{trenuno}
\end{equation}
\end{mathletters}
turn to be connected. All the fundamental fermions of the 
standard model have mass and are Dirac fermions, say $\psi$, in four states, 
$\psi_L$, $\psi_R$ and the charge conjugated states 
$(\psi_L)^c=(\psi^c)_R$ and $(\psi_R)^c=(\psi^c)_L$  corresponding to 
the antiparticles. To each Dirac fermion we define the charge conjugated 
field
\begin{equation}
\psi^c = C\bar\psi^{\sf T}
\label{tredue}
\end{equation}
which also satisfies the Dirac equation,
\begin{equation}
(i\hbar\gamma^\mu\partial_\mu - mc)\psi^c=0.
\label{trentre}
\end{equation}
The neutrinos are the only fundamental fermions that do not have electric 
charge~\cite{Footcharged}, 
even though they can have the character of Majorana fermions to 
which $\psi^c=({\rm phase\,\,factor})\psi$.
\par
Finally, in relation to the four fundamental interactions there is a 
curious coincidence. Why `4-3-2-1'? Why four dimensions of the space-time? 
Why SU(3) for the color group? Why SU(2) for the weak isospin group? 
Why U(1) for the quantum electrodynamics?
\section{Vacuum Catastrophe}
Let us now describe how the standard model triggers 
the `vacuum catastrophe'~\cite{Adleretal}. 
Consider a real scalar field $\phi$ whose dynamics is contained in the 
Lagrangian density
\begin{equation}
{\cal L}(\phi)=\frac{1}{2}\partial_\mu\phi\partial^\mu\phi-V(\phi)
\label{trquat}
\end{equation}
where the first term with derivatives is the kinetic one and the potential 
is
\begin{equation}
V(\phi)=-\frac{1}{2}m^2\phi^2+\frac{1}{4}\lambda\phi^4
\label{trncinq}
\end{equation}
which contains the quadratic mass term and the interaction term 
$\lambda\phi^4$. The Lagrangian ${\cal L}(\phi)$ is invariant under the 
discrete symmetry transformation 
${\cal L}(\phi)\rightarrow {\cal L}(-\phi)$, then 
${\cal L}(-\phi)={\cal L}(\phi)$. In the standard model of elementary 
particles the more general scalar potential which allows the implementation 
of the spontaneous gauge symmetry breaking 
\begin{equation}
{\cal G}_{321}                                                 
\rightarrow {\rm SU}(3)_c\times {\rm U}(1)_{\cal Q}
\label{tresei}
\end{equation}
can be written as
\begin{equation}
V(\Phi^\dagger\Phi)=a\Phi^\dagger\Phi+b(\Phi^\dagger\Phi)^2
\label{trnsette}
\end{equation}
where $a$ and $b$ are two numbers, 
$\Phi^\dagger\equiv (\Phi^*)^{\sf T} = (\Phi^{\sf T})^*$, 
and
\begin{equation}
\Phi=\left (
\begin{array}{c}
\phi^+\\
\phi^0
\end{array}
\right )\sim ({\bf 1},{\bf 2},Y=+1)
\label{trenotto}
\end{equation}
is the doublet of scalar fields in the SU(2) fundamental representation. 
In the process of spontaneous symmetry breaking only the neutral component 
obtains a vacuum expectation value
\begin{equation}
\langle 0|\phi^0|0\rangle\equiv\langle\phi^0\rangle_0 = 
\frac{1}{\sqrt 2}\left (
\begin{array}{c}
0\\
\sigma
\end{array}
\right )
\label{trnove}
\end{equation}
so that, in terms of $\sigma$, the potential becomes
\begin{equation}
V(\sigma)=\frac{a}{2}\sigma^2 + \frac{b}{4}\sigma^4
\label{quoranta}
\end{equation}
and defining
\begin{equation}
a\equiv -m^2,\quad b\equiv \lambda;\quad m>0
\label{quaruno}
\end{equation}
it results
\begin{equation}
V(\sigma)=-\frac{1}{2}m^2\sigma^2 + \frac{1}{4}\lambda\sigma^4.
\label{quardue}
\end{equation}
The minimum of this potential is determined by the general conditions
\begin{equation}
V^\prime\equiv\frac{\partial V}{\partial\sigma}=0,
\quad V^{\prime\prime}\equiv\frac{\partial^2V}{\partial\sigma^2}>0
\label{quatre}
\end{equation}
which in terms of the potential parameters $m$ and $\lambda$ are
\begin{mathletters}
\begin{equation}
V^\prime = -m^2\sigma + \lambda\sigma^3 = 0,
\label{quaquatr}
\end{equation}
and
\begin{equation}
V^{\prime\prime} = -m^2 + 3\lambda\sigma^2 > 0
\label{quarcinq}
\end{equation}
\end{mathletters}
so that the minimum of the potential
\begin{equation}
V(\sigma=\sigma_\pm) = -\frac{m^4}{4\lambda}
\label{quarsei}
\end{equation}
occurs exactly when
\begin{equation}
\sigma_\pm = \pm (m^2/\lambda)^\frac{1}{2}.
\label{quarsette}
\end{equation}
On account of the fact that $V(\sigma_+)=V(\sigma_-)$ either 
$V(\sigma_+)$ or $V(\sigma_-)$ are equivalent minimal values of the 
potential the symmetry of reflection $\phi\rightarrow-\phi$ in the Lagrangian 
${\cal L}(\phi)$ given in Eq.~(\ref{trquat}) is broken 
by the choice of one of the 
vacuum states. A symmetry of the Lagrangian that is not respected by the 
vacuum state is a spontaneously 
broken symmetry. In physics, the word `vacuum' not only has the meaning of 
`empty space'  but also denotes the fundamental state, the state of lower 
energy in a quantum field theory. In general, the vacuum state is a Lorentz 
invariant eigenstate. By choosing a unitary gauge, the doublet of scalar 
fields can be placed as
\begin{equation}
\Phi = \frac{1}{\sqrt 2}\left (
\begin{array}{c}
0 \\
\sigma + H
\end{array}
\right )
\label{quasette}
\end{equation}
where $H$ is the Hermitian field associated to the Higgs boson of the 
electroweak sector of the standard model. Arising out of this, in terms 
of $H$, the scalar potential in Eq.~(\ref{trnsette}) will contain even terms 
of fourth order,
\begin{equation}
V(H) = -\frac{m^4}{4\lambda} - m^2H^2 + \lambda\sigma H^3+
\frac{\lambda}{4}H^4
\label{quarotto}
\end{equation}
including the quadratic mass term, $\frac{1}{2}M^2_H H^2$, with 
$M^2_H = 2m^2$ but also the constant term $-m^4/(4\lambda)$ which is the 
nonvanishing term of the potential when evaluated in the minimum states 
$\phi = \sigma_\pm$. It was suggested for the first time by 
Zel'dovich~\cite{Zel68} that this self-energy of vacuum is interpreted as a 
cosmological constant
\begin{equation}
\Lambda=\frac{8\pi G}{c^4} V(\sigma_\pm)
\label{cinqta}
\end{equation}
in the modified Einstein equations
\begin{eqnarray}
R_{\mu\nu}-\frac{1}{2}R g_{\mu\nu} & = & \frac{8\pi G}{c^4}T_{\mu\nu}-
\Lambda g_{\mu\nu}\nonumber \\ 
& = & \frac{8\pi G}{c^4}(T_{\mu\nu} - V(\sigma_\pm) g_{\mu\nu})
\label{cinquno}
\end{eqnarray}
where 
$G=6.673(10)\times 10^{-11}\,\,{\rm kg}^{-1}\,\,{\rm m}^3\,\,{\rm s}^{-2}$ is 
the Newton--Cavendish constant being the fundamental universal constant 
associated to the gravitational interaction. 
These equations contain the term that involves the cosmological constant 
$\Lambda$ whose experimental bound is
\begin{equation}
G\Lambda \lesssim 10^{-122}.
\label{cnqdue}
\end{equation}
Everything in the Universe conspire extremely well to generate and to keep 
an exceptionally small numerical value. Any phase transition as the 
spontaneous 
gauge symmetry breaking is always accompained by a modification in the 
vacuum energy which is at least 50 orders of magnitude higher than this 
limit~\cite{Weinberg89}. The limit that is established to the value of the 
cosmological constant contained in the Einstein equations of 
general relativity is the result of this conspiration that institutes the 
number $10^{-122}$. This number is not zero; however, it is the smallest 
fundamental number of physics. 
In order to establish this bound it is necessary the 
use of {\it natural units}. 
It is established the 
numerical adimensional value 
$c=\hbar =1$ for the velocity of light in the vacuum ($c$) and for the 
reduced Planck constant ($\hbar\equiv h/2\pi$) which are two universal 
physical constants associated to the Maxwell electromagnetic theory and 
 the theory of special relativity ($c$), and to the quantum mechanics 
($\hbar$). The gravitational constant $G$ is the universal physical constant 
involved in the constraint $G\Lambda\lesssim 10^{-122}$ on the cosmological 
constant $\Lambda$. 
\par
Let us now establish the Planck scale through the combination of the $c$, 
$G$, and $\hbar$ universal constants. The Planck energy scale is 
$E_{\sf P}=(\hbar c^5/G)^\frac{1}{2}\simeq 1.2211\times 10^{19}$ GeV. 
However, Planck had the idea that the universal physical fundamental 
constants are enough to determine natural units of energy (GeV), but also 
scales of lenght, $l_{\sf P}=(\hbar G/c^3)^\frac{1}{2}\simeq 
1.6161\times 10^{-35}\,\,{\rm m}$ and of time, 
$t_{\sf P}=(\hbar G/c^5)^\frac{1}{3}\simeq 5.3904
\times 10^{-44}\,\,{\rm sec}$.
\par
The high energy physics studies the quantum and relativistic regimes of 
nature. We always try to seek events among elementary particles which 
involves velocities near to that of light in vacuum $c$, and 
actions or angular momenta 
of the order of the reduced Planck constant, $\hbar$. This is the reason 
why it is tacitly placed $c=\hbar = 1$ which makes the high energy physics 
unidimensional so that it is possible to have only one fundamental 
magnitude. To the c.g.s. or m.k.s. units the fundamental dimensions are mass 
$[M]$, length $[L]$, and time $[T]$. In the natural units system the new 
fundamental dimensions are mass $[M]$, action $[S]$, and velocity $[V]$. 
In terms of natural units the lenghts and time are
\begin{equation}
[L]=\frac{[S]}{[M][V]},\quad [T]=\frac{[S]}{[M][V]^2}
\label{extuno}
\end{equation}
and being $p$, $q$, $r$ real numbers the general dimensional 
relation 
\begin{equation}
[M]^p [L]^q [T]^r = [M]^{p-q-r} [S]^{q+r} [V]^{-q-2r}
\label{extdue}
\end{equation}
in the c.g.s. or m.k.s. units has the natural units mass dimension $[M]^n$ 
with $n = p-q-r$. Thereby, for action and velocity we have $p=1$, $q=2$, 
$r=-1$, and $p=0$, $q=1$, $r=-1$ respectively, with $n=0$ for both of such 
magnitudes. The mass or energy, $E= M c^2$, length, and time have the 
$p=1$, $q=r=0$; $q=1$, $p=r=0$; and $r=1$, $p=q=0$ attributions with the 
respective natural units dimensions $n=+1,-1,-1$.  
In natural units, the energy, mass, and linear momentum 
have the same dimension, say measured in eV energy unit meanwhile time 
and distances have the 
inverted dimension (eV$^{-1}$). What is intended is to know how the world 
is at distances even shorter or at even higher energies~\cite{WH99}. 
This makes the present particle accelerators become the microscope solving 
today lenghts of $1.9733\times 10^{-19}\,\,{\rm meters}$, exactly equivalent 
to the energy scale of 
$1\,\,{\rm TeV} = 10^3\,\,{\rm GeV} = 10^{12}\,\,{\rm eV}$. 
The {\it detector physics}, by its turn, as the 
{\it neutrino physics}~\cite{SupK} and the 
{\it physics of cosmic rays}~\cite{Auger,Ginzburg} 
are revealing physics from below the ${\rm eV}$ scale up to energies even very 
higher than the ${\rm TeV}$ scale. 
\par
The Planck energy $E_{\sf P}$ is associated to the universal gravitation 
constant $G$ according to
\begin{equation}
\frac{1}{\sqrt G} = E_{\sf P}\approx 10^{19}\,\,{\rm GeV}
\label{cquure}
\end{equation}
and numerically,
\begin{equation}
G\approx 5.9\times 10^{-39} M_{{\rm p}^+}
\label{cququau}
\end{equation}
where $M_{{\rm p}^+}$ is the proton mass. Then, we have the following 
bound on the cosmological constant,
\begin{equation}
\Lambda\lesssim \frac{10^{-122}}{G}\approx 10^{-84} M^2_{{\rm p}^+}
\simeq 10^{-84}\,\,{\rm GeV}^2
\label{ciquacio}
\end{equation}
but the same order of magnitude is obtained by using the Planck scale of 
energy
\begin{equation}
\Lambda\lesssim 10^{-122}E^2_{\sf P}\simeq 10^{-84}\,\,{\rm GeV}^2.
\label{cquacdue}
\end{equation}
The limits realized in the process of desacelleration of the Universe 
imply
\begin{equation}
|\Lambda|\lesssim 4\times 10^{-84}\,\,{\rm GeV}^2\simeq 
10^{-52}\,\,{\rm m}^{-2}
\label{cqctre}
\end{equation}
on the value  of the cosmological constant which in fact coincides with 
the adimensional quantity $G\Lambda\lesssim 10^{-122}$.
\par
Let us now concentrate in the determination of the 
numerical value~\cite{IStewart} of the 
cosmological constant $\Lambda_{\rm SM}$ induced in the spontaneous gauge 
symmetry breaking ${\rm SU}(2)\times {\rm U}(1)^\prime\rightarrow {\rm U}(1)$ 
of the electroweak standard model. Recovering the Zel'dovich relation given 
in Eq.~(\ref{cinqta}) in which according to Eq.~(\ref{quarsei}) the 
potential acquires the value given by $-m^4/(4\lambda)$ we obtain
\begin{equation}
\Lambda_{\rm SM} = -\frac{2\pi G}{c^4}m^2\sigma^2_\pm
\label{cqcquattro}
\end{equation}
and by using the relation
\begin{equation}
\frac{G_{\rm F}}{\sqrt 2} = \frac{1}{2\sigma^2}
\label{cqaquattro}
\end{equation}
where $G_{\rm F} \simeq 1.02\times 10^{-5} M^{-2}_{{\rm p}^+}\simeq 
1.17\times 10^{-5}\,\,{\rm GeV}^{-2}$ is the Fermi universal constant of 
weak interaction, it results that the cosmological constant is
\begin{equation}
\Lambda_{\rm SM} = -\frac{\pi G}{2{\sqrt 2}c^4G_{\rm F}}M^2_H\approx 
-10^{-33}M^2_H
\label{cqcinque}
\end{equation}
being proportional to the square of the mass $M^2_H$ of the Higgs scalar 
boson. Hence, taking into account the experimental inferior limit 
$M_H>100\,\,{\rm GeV}$ it results
\begin{mathletters}
\begin{equation}
\Lambda_{\rm SM}\sim 10^{54}\Lambda_{\rm obs},\quad 
\label{cqsei}
\end{equation}
with
\begin{equation}
\Lambda_{\rm obs} = 10^{-122}/G\simeq 10^{-84}\,\,{\rm GeV}^2
\label{cqselinea}
\end{equation}
\end{mathletters}
comprehending a numerical factor representing the major discordance between 
theory and experiment of the whole physics. The standard model containing 
the description of the non-gravitational interactions quantum dynamics and 
is extremely successful in its concordance with the experiments presents 
this problem. Every time that the Higgs mechanism~\cite{Higgsssb,Kibble} 
operates 
in order to generate masses in a gauge theory it is generated besides a 
cosmological constant. This is a problem not only belonging to the standard 
model, but also to the grand unification theories (GUTs) and supersymmetric 
(SUSY) schemes.
\par
Connected to all these theories, it always appears a hierarchy problem: the 
energy of the vacuum state of the Universe is zero with a precision of 
$10^{-122}$. In the dynamics of vacuum, where a cosmological phase transition 
is induced by a scalar field, when we consider the field as a 
temperature-dependent function, there is a reciprocal mechanism giving a 
vanishing cosmological constant. The scalar sector of the theory has a 
vanishing vacuum energy that is induced by itself. This could have occurred 
in the ocasion of a cosmological phase transition in which the Higgs 
potential is corrected accordingly~\cite{HwkgLHSS}
\begin{equation}
{\cal V}(\phi) = - \mu^2(T) + \lambda\phi^4
\label{sesstdue}
\end{equation}
being the radiative corrections until first order ${\cal O}(\lambda)$ in 
the perturbative serie. The mass parameter in terms of temperature is
\begin{equation}
\mu^2(T)=\mu^2_0 - \kappa T^2
\label{sesantre}
\end{equation}
where the coefficient $\kappa$ depends either on $\phi$ be representing a 
gauge singlet or any multiplet in the fundamental or adjoint gauge group 
representations. In general, $\kappa\sim\lambda$. The phase transition occurs 
when the temperature $T$ is lower than the critical temperature,
\begin{equation}
T_C = (\mu^2_0/\kappa)^\frac{1}{2},\quad T < T_C\,,
\label{sesnquat}
\end{equation}
when a new state of minimum is created in $\phi = \sigma$, which is the 
vacuum state. In the new phase, the cosmological constant is determined by
\begin{equation}
{\cal V}({\phi}) = \frac{\mu^4}{4\lambda}=
\frac{(\mu^2_0 - \kappa T^2)^2}{4\lambda}
\label{sestcinqu}
\end{equation}
which naturally depends on temperature, $T$. The idea that a cosmological 
constant depends on the temperature is not a new one~\cite{Beciu}. 
Nonetheless, the possibility of a vacuum state decaying with temperature 
throughout the whole history of the Universe~\cite{Ozeretal} is severally 
suppressed not to interfere with results of the phase of 
nucleosynthesis~\cite{Freese}. 
It is possible to build up finite temperature field theories~\cite{DJackiw} 
when the dependence of the cosmological 
constant with temperature is limited only to phase transitions. 
\section{The Baum--Hawking--Coleman Solution}
An attempt to solve the vacuum problem was proposed by E. Baum~\cite{Baum} 
in 1983, S. W. Hawking~\cite{Hawk84} in 1984 and S. Coleman~\cite{Coleman} 
in 1988. The BHC solution has two crucial ingredients:\\
(i) The observable value of the cosmological constant $\Lambda$ is not 
absolutely a 
c-number fundamental parameter. On the contrary, it is necessary to consider 
Universes with different values of $\Lambda$ distributed according to a 
certain distribution, ${\cal P}(\Lambda)$, in which $\Lambda$ is promoted to 
quantum dynamical variable, a q-number.\\
This presumption was successful by studies of topological solutions such 
as {\it baby universes} and {\it wormholes}~\cite{HLGA} which finally led 
to the fact that 
all coupling constants, including $\Lambda$ 
are dynamical variables.\\
(ii) It is supposed that the probability distribution 
${\cal P}(\Lambda)$ is possible to be determined. Baum~\cite{Baum} and 
Hawking~\cite{Hawk84} have proposed that
\begin{equation}
{\cal P}(\Lambda)\propto \exp\{-S(\Lambda)\}
\label{sessantsei}
\end{equation}
where the action is
\begin{equation}
S(\Lambda) = -3\pi\frac{M_{\sf P}}{\Lambda}
\label{sesstsette}
\end{equation}
and thus ${\cal P}(\Lambda)$ is strongly pronounced in $\Lambda =0$, 
\begin{equation}
\lim_{\Lambda\rightarrow 0} {\cal P}(\Lambda) = \infty.
\label{sesttextr}
\end{equation}
Coleman~\cite{Coleman} proposed
\begin{eqnarray}
{\cal P}(\Lambda) & \propto & \exp\{\exp\{-S(\Lambda)\}\}\\
& = & \exp\{\exp\{3\pi\,M_{\sf P}/\Lambda\}\}
\label{ssstotto}
\end{eqnarray}
which has a peak exceptionally more accentuated when $\Lambda =0$. \\
However, it is not clear until which point the proposition of solution by 
Baum--Hawking--Coleman~\cite{Baum,Hawk84,Coleman} for the cosmological 
constant is efficient~\cite{Lavre91}, considering that, for instance, 
the results of ${\cal P}(\Lambda)$ can not be entirely based on the Euclidean 
functional integration of the quantum gravity. The modified Einstein 
equations [Eq.~(\ref{cinquno})] follow from the action
\begin{equation}
S(\Lambda) = (16\pi\, G)^{-1}\int {\rm d}x\,{\sqrt g}\,(2\Lambda R)
\label{sesanove}
\end{equation}
which is the most general form of a local action consistent with the 
invariance principles of general relativity. The BHC solution appeals to 
the controversial ways of quantum gravity Euclideanization~\cite{Gibbetal} 
as well as the action $S(\Lambda)$ is related to the possible non-localities 
as a result of non-perturbative topological effects as the 
{\it wormholes} able to  connect two distinct plane Universes or two 
regions of the same Universe. 
\section{Baryon Asymmetry of the Universe}
One of the fundamental questions of quantum particle cosmology concerns to the 
mechanisms which could have generated the baryon asymmetry of the Universe, 
that is the observed disequilibrium between matter and antimatter. The ratio 
between the densities of the number of baryons, $n_b$, and of photons, 
$n_\gamma$, is given by the estimatives of the parameter~\cite{Boesg85}
\begin{equation}
\eta\equiv\frac{n_b}{n_\gamma}\simeq 10^{-10\pm 1}
\label{setanta}
\end{equation}
where
\begin{mathletters}
\begin{equation}n_b\simeq 10^{-5}\,\,{\rm cm}^{-3}
\label{setantuno}
\end{equation}
and
\begin{equation}
n_{\gamma}\simeq 400\left (\frac{T}{T_0}\right )^3\,\,{\rm cm}^{-3}.
\label{setandue}
\end{equation}
\end{mathletters}
The data of the COBE-FIRAS satellite~\cite{Turner} indicate a temperature 
$T_0 = (2.726\pm 0.001)\,\,{\rm K}$ of the cosmic background radiation. The 
cosmic background radiation spectrum adjusts to a thermal bath spectrum with 
incredible precision. For $T=T_0$, the density of the number of photons is 
$400\,\,{\rm cm}^{-3}$.\\
The asymmetry between baryons and antibaryons can be justified in the 
following manner: the Universe started with a complete symmetry between 
matter and antimatter in the standard description of the Big Bang and in the 
subsequent evolution it was generated a baryon number. In principle, it 
is possible provided that three conditions be observed:
\begin{enumerate}
\item There is a kind of interaction which violates the conservation of the 
      baryon charge at fundamental level;
\item There are interactions which violates the symmetry of charge 
      conjugation $C$ and the $CP$ combined symmetry of charge conjugation 
      and parity so that they induce asymmetry among processes involving 
      particles and antiparticles;
\item There are shifts in the thermal equilibrium state of the scalar and 
      vector particles which mediate interactions that violate the 
      conservation of baryon number.
\end{enumerate}
Strictly in the realm of the standard model of the non-gravitational 
interactions it is not possible to generate the violation of the 
conservation of the baryon number. Nevertheless, there are 
instanton~\cite{GtHooft} kind solutions of the corresponding equations 
of motion. It is associated to the instantons a quantum number, the 
topological charge, and induces interactions between quarks and leptons as
\begin{equation}
(u+u+d) + (c+c+s) + (t+t+b)\rightarrow e^+ + \mu^+ + \tau^+
\label{setntre}
\end{equation}
in which there are three quarks comprehended between parentheses with 
different SU(3)$_c$ color charges but that are elements of the same 
doublet of weak isospin. However, such transitions are suppressed by a 
factor $\exp\{-4\pi/\alpha_{\rm W}\}$ where 
$\alpha_{\rm W}=\alpha/\sin^2\theta_{\rm W}$ and
\begin{equation}
\alpha = \frac{e^2}{4\pi\epsilon_0\hbar c}=
\frac{1}{137.035\,989\,5(61)}
\label{setanquatr}
\end{equation}
is the fine structure constant and 
$e=1.602\,177\,33(49)\times 10^{-19}\,\,{\rm C}$ is the unit of electric 
charge. Therefore,
\begin{equation}
\exp\{-4\pi/\alpha_{\rm W}\}\simeq\exp\{-398\}\simeq 10^{-172}
\label{setncinque}
\end{equation}
which is $50$ orders of magnitude even smaller than the bound on the 
cosmological constant, $G\Lambda\lesssim 10^{-122}$ (!).
\section{Casimir Effect}
There are phenomena of quantum nature~\cite{Mohrhoff} 
related to the presence of fluctuations 
in the vacuum state of the quantized fields. One of the recognizable 
predictions of the quantum electrodynamics (QED) was announced in 1948 by 
Hendrik B. Casimir~\cite{HBCasimir} who proposed the effect where a variation 
of the vacuum energy density produce an attractive force between two plane 
and electric conductor plates. For two metallic parallel plates, with area 
$A$ and separated by a distance $d$, the attractive Casimir force, $F(d)$, by 
unit of area, $A$, is~\cite{Milloni}
\begin{equation}
\frac{F(d)}{A} = \frac{\pi^2}{240}\frac{\hbar c}{d^4}\propto\frac{1}{d^4}.
\label{setntasei}
\end{equation}
This attraction was experimentally confirmed, for the first time, after 
ten years~\cite{Sparn}. For plates of $1\,\,{\rm cm}^2$ of area and being 
$d = 0.5\,\,\mu{\rm m}$, the measured value of the force was 
$\sim 2\times  10^{-6}\,\,{\rm N}$ that, in fact, agrees with 
Eq.~(\ref{setntasei}). Recently, the effect of the Casimir force was shown in 
the scale between $0.6$ and $6\,\,\mu{\rm m}$. The pressure of the vacuum 
state between two very near conductor surfaces is considered conclusively 
demostrated~\cite{LamBut} due to the fluctuations of the fundamental quantum 
state. \par
This kind of force is part of a group of effects due to variations of the 
energy density of the vacuum state now collectively called Casimir 
forces~\cite{MilShih}. The Lamb~\cite{Lamb} shift verified in 1947 is a 
result of the vacuum energy shift of the hydrogen atom. Hence, the 
quantum vacuum state is much more than empty space.
\section{Anthropic Principle}
The anthropic principle~\cite{Cartetal} states that the Universe is the 
way it is because otherwise there would not be anyone to ask why it is this 
way. Many of the aspects of the Universe would be determined by the condition 
that there is intelligent life in it. One of these aspects~\cite{WheDic} is 
referent to the dimensions of the Universe since it was smaller, with a 
larger matter density, it would already had suffered a gravitational collapse 
before intelligent life had evoluted. Another aspect refers to the lifetime 
of the proton whose experimental 
bound is larger than $10^{16}\,\,{\rm years}$ because 
on the contrary living beings would not survive to the effects of the 
ionizing particles produced by the proton decay in their organisms. There 
is almost one mol of stars in the Universe. How many of them there had 
all the conditions for life to become intelligent? A lot of them? 
May be. Besides, it is possible that `there is something unique relatively 
to the Man and to the planet in which he lives' according to the geneticist 
T.G. Dobzhansky (1900--1975). 
\par
The anthropic principle~\cite{Cartetal} in a version a lot stronger states 
that the natural laws are complete only in the case that there is intelligent 
life once quantum mechanics would not have sense without the observer. 
Weinberg~\cite{Weinberg89} however in relation to this strong version 
states that `despite the fact that science is clearly impossible without 
scientists it is not clear that the Universe is impossible without science.'
\\
Finally, the weak anthropic principle searchs for an explanation of 
what are the 
possible eras and parts of the Universe where we could be calculating 
which eras and parts of the Universe we can live and illustrates the first 
use of anthropic arguments in modern physics with the Dicke~\cite{WheDic} 
solution to the problem proposed by Dirac. Still in $1937$ 
Dirac~\cite{PDirac1937} realized the combination with dimension of time of 
physical universal constants providing the age of the Universe,
\begin{equation}
t_U = \frac{\hbar}{G\,c}\,
\frac{\alpha}{m^2_{{\rm e}^-}\,m^{\mbox{}}_{{\rm p}^+}}
\simeq 4.5\times 10^{10}\,\,{\rm years}.
\label{settntsette}
\end{equation}
Let us observe that the time scale goes from the 
Planck scale
\begin{equation}
t_{\sf P} = (\hbar G/c^5)^\frac{1}{2}\simeq 10^{-44}\,\,{\rm sec}
\label{setntotto}
\end{equation}
up to the scale determined by the Hubble~\cite{Hubblecon} constant, $H_0$, the 
fundamental constant of cosmology, which establishes the age of the Universe,
\begin{equation}
t_U = H_0^{-1}\simeq 10^{10}\,\,{\rm years}\simeq 10^{17}\,\,{\rm sec}
\label{settnove}
\end{equation}
so that the time scale has the bounds
\begin{equation}
t_{\sf P}\simeq 10^{-44}\,\,{\rm s}\leq t\leq t_U\simeq 10^{17}\,\,{\rm s}
\label{ottantah}
\end{equation}
determined by the Planck and Hubble scales, respectively. According to the 
modified Einstein equations, Eq.~(\ref{cinquno}), involving the 
$\Lambda g_{\mu\nu}$ term of the cosmological constant, the universal 
expansion law of the scale factor $R(t)$ of the Universe is given by
\begin{equation}
H^2\equiv \left (\frac{\dot{R}}{R}\right )^2 = 
\frac{8\pi G}{3}\rho_{\rm M}+\frac{\Lambda}{3}-\frac{k}{R^2},
\label{ottantuno}
\end{equation}
where ${\dot{R}}={\rm d}R(t)/{\rm d}t$, $\rho_{\rm M}$ is the matter density 
of the Universe; $k=-1,0,+1$ to Universes that are curved 
open, flat, and curved closed and, finally, $H$ is the 
Hubble constant whose present value is
\begin{equation}
H_0 = 100\,h_0\,\,{\rm km}\,\,{\rm s}^{-1}\,\,{\rm Mpc}^{-1}
\label{ottndue}
\end{equation}
with $1\,{\rm Mpc}\,\,({\rm megaparsec})\simeq 3.1\times 10^{19}\,\,{\rm km}
\simeq 3$ 
light-year. In a natural unit, 
\begin{equation}
H_0 = 2.13\,h_0\times 10^{-42}\,\,{\rm GeV}
\label{ottdlinha}
\end{equation}      
with an adimensional ignorance parameter in the interval $0.5 < h_0 < 0.87$. 
The Supernova Cosmology Project data~\cite{SCP} gives the value
\begin{equation}
H_0 = (63.1\pm 3.4\pm 2.9)\,\,{\rm km}\,\,{\rm s}^{-1}\,\,{\rm Mpc}^{-1}
\label{otndulinha}
\end{equation}
and a theoretical deduction~\cite{Marcinkowski} gives
\begin{equation}
H_0 = 61.4\,\,{\rm km}\,\,{\rm s}^{-1}\,\,{\rm Mpc}^{-1}.
\label{lacrg}
\end{equation}
The Hubble law above contains three terms determining the expansion of the 
Universe. 
The first one is the common term of matter, the second one is that 
of the cosmological constant and, finally, the last one is the space-time 
curvature term. 
\par
The Dicke solution to the Dirac problem indicates that the question of the 
age of the Universe can appear only when the conditions for the existence 
of life are appropriate and correct. The Universe should have enough age to 
some stars having completed their permanence in the principal sequence and 
produced heavy chemical elements necessary to the existence and maintenance 
of life. Otherwise, other stars should be young enough so that they are still 
producing energy, i.e., light and heat for life, by means of thermonuclear 
reactions.
\par
Dirac argumented that if the connection established in 
Eq.~({\ref{settntsette}) were anyhow fundamental as the age of the 
Universe increases linearly with the time $t$, at least one constant on the 
Eq.~(\ref{settntsette}) should vary with the time and conjectured that the 
universal constant $G$ of gravitation varies according to $1/t$. In $1985$ 
Zee~\cite{Zee} applied the same procedure to the cosmological constant based 
in the condition $G\Lambda\lesssim 10^{-122}$ so that, if 
$G=G(t)\propto 1/t$, 
then $\Lambda (t)\propto t$. According to Weinberg, it is perfectly 
reasonable to apply anthropic considerations in order to know in which era 
or part of the Universe we could exist and, therefore, what values of the 
cosmological constant we could observe.
\section{Conclusions}
Although Einstein called his attempt of introducing a cosmological constant 
in their equations of general relativity 
to keep the Universe static the worst mistake of his whole 
life~\cite{Gamow}, he did not suspect that a so large 
cosmological constant would be induced in 
any theory with spontaneous gauge symmetry breaking. The cosmological 
constant remains an essential quantity, either of cosmology or of 
high energy physics~\cite{Bernstein}. In the case of cosmology due to 
the strong constraint 
$G\Lambda\lesssim 10^{-122}$ and for the high energy physics, 
in the context of quantum field theories with local gauge symmetries as the 
standard model of non-gravitational interactions, to a cosmological constant 
corresponding to the energy density associated to the vacuum state in the 
process of spontaneous gauge symmetry breaking which is not 
canceled. The vacuum is a busy place in any quantum theory and should 
gravitate~\cite{Polch99,Polch96,Coll97}.
\par
The seriousness of the vacuum state energy problem~\cite{Vafa} 
conduced a lot of physicists 
to believe that it should have exist some mechanism of cancellation or that 
quantum considerations of cosmology would justify a vanishing value. 
The problem is 
that it is not possible to find any symmetry that warrant an identically zero 
value~\cite{SCol88} and arguments of quantum cosmology are still based in the 
quantum Euclidean gravity. Even the Peccei--Quinn~\cite{PeccQuinn} symmetry 
results in an imperfect cancellation. A cancellation mechanism is 
viable~\cite{PiTon98} 
in the leptoquark-bilepton flavor dynamics~\cite{PPF}. There is also the 
possibility of a probability distribution in which the cosmological 
constant takes different values in cosmological theories with a large 
number of sub-Universes with different terms in the wave function of the 
Universe~\cite{Weinb96}.
\par 
Finally, it is also considered the intriguing possibility of the 
energy content of the vacuum state is zero but, in the moment, we would be 
in a phase transition in which the Universe is in the state of false 
vacuum~\cite{Hill}. 
The energy scale of this transition corresponds to 
$(10^{-46}\,\,{\rm GeV}^4)^\frac{1}{4}\simeq 3\times 10^{-3}\,\,{\rm eV}$ 
which can be near to the value of lightest neutrino mass, which would also 
solve the solar neutrino problem~\cite{ConchaPRL}.
\acknowledgments 
N.O.R. wishes to thank the CNPq (Brazil) for financial support.

\end{document}